\documentstyle[12pt]{article}

\def\be{\begin{equation}}
\def\ee{\end{equation}}
\def\bea{\begin{eqnarray}}
\def\eea{\end{eqnarray}}
\def\cO{{\cal O}}
\def\GeV{{\rm GeV}}
\def\mbar{{\overline{m}}}
\def\simlt{\stackrel{<}{{}_\sim}}
\def\simgt{\stackrel{>}{{}_\sim}}

\begin{document}
\hbadness=10000
\hbadness=10000
\begin{titlepage}
\nopagebreak
\def\thefootnote{\fnsymbol{footnote}}
\begin{flushright}
{\normalsize
BONN--TH--98--11\\
SFB--375/297, TUM--TH--314/98\\
TU--547, RCNS--98--07\\
May 1998}\\
\end{flushright}
\vspace{0.5cm}
\begin{center}

{\large \bf   Relic Abundance of Neutralinos
in Heterotic String\\ Theory: 
Weak Coupling vs. Strong Coupling}

\vspace{0.8cm}

{ Y.~Kawamura$^{1}$\footnote{On leave of absence from Department of
Physics, 
Shinshu University, Japan. Humboldt Fellow}}, 
{ H. P.~Nilles$^{1,2}$},
{ M.~Olechowski$^{3}$\footnote{
On leave of absence from Institute of Theoretical Physics, 
Warsaw University, Poland.}}\\
and \\
{ M.~Yamaguchi$^{4}$}

\vspace{0.5cm}
{\sl $^1$ Physikalisches Institut, Universit\"at Bonn \\
   D-53115 Bonn, Germany, \\
$^2$ Max-Planck-Institut f\"ur Physik\\
   D-80805 M\"unchen, Germany\\
$^3$ Physik Department, Technische Universit\"at M\"unchen\\
   D-85747 Garching, Germany\\
$^4$ Department of Physics, Tohoku University \\
     Sendai 980-8578, Japan\\}

\end{center}
\vspace{0.5cm}

\nopagebreak

\begin{abstract}
The relic abundance of stable neutralinos is investigated in 
$E_8 \times E_8'$ heterotic string theory when supersymmetry is 
spontaneously broken by hidden-sector gaugino condensates.
In the weak coupling regime, very large scalar masses (compared to gaugino 
masses) are shown to lead to a too large relic abundance of the neutralinos, 
incompatible with cosmological observations in most of parameter space.
The problem does not arise in the strong coupling regime (heterotic
M--theory) because there scalar and gaugino masses are generically of the
same order of magnitude. 
\end{abstract}
\vfill
\end{titlepage}
\pagestyle{plain}
\newpage

\def\thefootnote{\arabic{footnote}}
\setcounter{footnote}{0}

%
%
\section{Introduction}

Although the $E_8 \times E_8'$ heterotic string theory has been an
attractive candidate for a unified theory including gravity, its weak
coupling regime seems to suffer from some phenomenological drawbacks. 
One of them is that the string unification scale is more than one order 
of magnitude higher than the GUT scale of about $3 \cdot 10^{16}$ GeV. 
This makes the
picture of the gauge coupling unification in this framework rather
complicated.  Another possible problem arises when one considers
supersymmetry (SUSY) breaking via gaugino condensates in the hidden
$E_8'$ sector 
(which is so far the most compelling mechanism of supersymmetry breaking) 
and looks into the structure of soft masses
\cite{DIN,DRSW,IN}. It was shown that gaugino masses are much smaller
than scalar masses. This hierarchical structure among the soft masses
may cause phenomenological and/or cosmological problems.

Recent developments of string theories make it possible to analyze their
strong coupling regime. In particular, we
now know that the strongly coupled  $E_8 \times E_8'$
heterotic string is described by the M-theory compactified on
$S^1/Z_2$ \cite{HW}.  Concerning the first problem mentioned 
above (on the discrepancy of
the scales), the M-theory description gives a simple solution. 
Namely, by adjusting the length of the interval $S^1/Z_2$, one can 
get the correct value of the Planck mass. The GUT scale 
(which can be identified with 
the compactification scale of a six dimensional Calabi-Yau manifold)
 is only by a factor of about 2 smaller than the fundamental mass 
scale in the theory \cite{C-unif,BD}.

Concerning the question of the supersymmetry breaking in the gaugino
condensation scenario, detailed analyses were recently worked out
\cite{Horava,NOY,LOW,LT}. 
(for related work in a somewhat different context see 
\cite{AQ,LLN,DG,Dudas,CKM,MP}).
It turns out, in the strong coupling regime, 
that the hierarchy among the soft masses disappears and gauginos and
scalars are generically in the same mass range which is assumed to be
at the electroweak scale.

The purpose of this paper is to make a comparison of phenomenological
and cosmological consequences between the weak and strong coupling
regimes of the heterotic string theory with supersymmetry broken by
the hidden-sector gaugino condensate.  Among other things, the question
of the relic abundance of the lightest supersymmetric particle (LSP)
highlights the difference between the two cases and thus we shall
focus on this issue in the present paper.  One expects that large masses of
the scalars in the weak coupling case may suppress the annihilation rates 
of the neutralinos, resulting in too large relic abundance which is in
contradiction with cosmological observations. We will closely study
the relic abundance in this regime and show that this is indeed the
case in most of the parameter space. On the other hand, we will point
out that the strong coupling regime does not encounter this
overclosure problem. Throughout this paper,
we assume that the low-energy effective theory is the supersymmetric
standard model with the minimal particle content (MSSM).

The paper is organized as follows. In the next section, we review SUSY
breakdown via gaugino condensation and its consequences for the soft SUSY
breaking parameters in the framework of the weakly coupled heterotic
string theory.  In section 3, we investigate the SUSY spectrum at the
electroweak scale based on the renormalization group equations (RGEs)
in the MSSM and the radiative electroweak breaking scenario.  In
section 4, we study in some detail the relic abundance of the
neutralinos in the weak coupling case and show that in the most of the
parameter space, the neutralino abundance is too large,  in
contradiction with cosmological observations.  Then we turn to the
case of the strong coupling regime in the subsequent section. Section 6 is
devoted to conclusions.

%
%
\section{Weakly coupled heterotic string theory}

Let us first review the soft SUSY breaking terms derived from the 
10-dimen\-sional weakly coupled heterotic string theory with 
$E_8 \times E_8'$ gauge group.
For simplicity, we discuss the 4-dimensional effective model 
with $E_6 \times E_8'$ gauge group
obtained through the dimensional reduction with the standard embedding.
Then the K\"ahler potential is given by \cite{Witten,DIN}
\begin{eqnarray}
G = -\log(S+\bar{S})-3\log(T+\bar{T}-2|C_i|^2)+\log|W(C)|^2
\label{G}
\end{eqnarray}
where $S$, $T$ and $C_i$ are the dilaton, the overall modulus 
and the matter fields, respectively.
The superpotential $W(C)$ is given by
\begin{eqnarray}
W(C) =  d_{ijk}C_i C_j C_k .
\label{W}
\end{eqnarray}
The gauge kinetic functions of $E_6$ and $E_8'$ are given by \cite{DIN,IN}
\begin{eqnarray}
f_6 =  S+\epsilon T ,~~~~~~ f_8 =  S-\epsilon T ,
\label{f}
\end{eqnarray}
respectively.  Here the terms involving $T$  originate from
the one loop corrections, which are related to the Green-Schwarz anomaly
cancellation counter terms, and $\epsilon$ is a  small
parameter. This result (\ref{f}) is not affected in higher 
orders of the perturbative expansion in the string coupling constant,
since there exist no higher loop corrections beyond one loop
\cite{SV,Nilles}.

We assume that the hidden $E_8'$ gaugino $\lambda$ 
(or the gaugino of a smaller gauge group $H'$ obtained from $E_8'$ e.g.\ 
through the Wilson line mechanism) condenses 
\begin{eqnarray}
\langle \lambda \lambda \rangle =  \Lambda^3, 
\label{gaugino-condensation}
\end{eqnarray}
where $\Lambda$ is the energy scale at which the gauge coupling 
of the gauge group $E_8'$ (or $H'$) becomes large.
The gaugino condensation can trigger supersymmetry breaking
as we infer from the expression for the $F$-components of the chiral
supermultiplets \cite{FGN}
\begin{eqnarray}
F_I =  (G^{-1})_I^J (\exp(G/2) G_J + {1 \over 4}f_J (\lambda \lambda)) +
... 
\label{FI}
\end{eqnarray}
where the indices $I$ and $J$ run over all chiral multiplets:  
$\Phi_I = (S, T, C_i)$.
We find that SUSY is broken by the $F$-term of the overall modulus 
field, i.e., $\langle F_S \rangle = \langle F_i \rangle =0$ 
and $\langle F_T \rangle \neq 0$,
and the vacuum energy vanishes in this approximation. 
Then the gravitino mass is given by
\begin{eqnarray}
m_{3/2} = {\langle F_T \rangle \over \langle T+\bar{T} \rangle}
\sim {\Lambda^3 \over m_{Pl}^2} 
\label{m3/2}
\end{eqnarray}
where $m_{Pl}$ is the Planck mass.  We can calculate the soft SUSY
breaking terms in the observable sector using the functions (\ref{G}),
(\ref{W}) and (\ref{f}).  The no-scale structure observed in (\ref{G})
\cite{no-scale} yields vanishing scalar masses, which appears as
a consequence of the 
assumed simplified nature of compactification.
In more general terms it is valid  only at the classical level,
and there only
for fields with modular weight $-1$ under $T$-duality \cite{BIM}. A
matter field which has modular weight other than $-1$ will have a
different  K\"ahler potential.  Furthermore, the K\"ahler potential
for all fields will, in general, receive sizable radiative
corrections. Thus, we expect the magnitude of the scalar
masses to be
\begin{equation}
 m_i =\cO(m_{3/2}) \label{scalar-mass}
\end{equation}
rather than exactly zero. The detailed structure of the scalar mass spectrum
is strongly model-dependent. 
For the gaugino masses, we find a situation that is simpler, e.g.
the mass of the gaugino of the $E_6$ gauge group, $M_{1/2}$, is given by
\begin{eqnarray}
M_{1/2} &=& {\langle f_6^I F_I \rangle \over \langle f_6+\bar{f}_6
\rangle}
= {\epsilon \langle F_T \rangle \over \langle f_6+\bar{f}_6 \rangle}
\label{M1/2}
\end{eqnarray}
where $f_6^I$ is the derivative of $f_6$ with respect to 
$\Phi_I$ and the relations, 
$\langle F_S \rangle = 0$ and $\langle F_T \rangle \neq 0$, 
have been used.
The magnitude of $M_{1/2}$ is thus estimated as 
\begin{equation}
M_{1/2}= \cO(\epsilon m_{3/2})\label{gaugino-mass}
\end{equation}
as far as $\langle S \rangle$ and $\langle T\rangle$ are of $\cO(m_{Pl})$. 
Hence, we find that the gaugino mass is much smaller than
the scalar masses, i.e., $|M_{1/2}|=\cO({\epsilon})|m_i|$, with
$\epsilon$ of the order of $10^{-2}$ or even less. 
The same applies to the masses of the gauginos present in the MSSM after $E_6$ 
is broken to the standard model gauge group.

%
%
\section{Soft SUSY breaking spectrum at low energies}

We consider models in which the soft scalar masses are much bigger 
than the soft gaugino masses at high energies of the order of $M_X$ 
(the GUT scale or the string scale) and want to calculate 
 the relic abundance of the LSP. 
To do this we need information about the soft SUSY breaking 
terms at low energies of the order of the weak scale $M_Z$. 
We assume that the observable gauge group $E_6$ is broken down to 
the standard model gauge group $SU(3)_C \times SU(2)_L \times U(1)_Y$ 
at the high energy scale $M_X$. The model below that scale is the MSSM.
The RGEs of the MSSM and the assumption 
about radiative breakdown of the electroweak symmetry will be used 
to get information about the low energy soft terms. 

Let us first estimate the order of magnitude of the high energy 
gaugino masses. The experimental bounds on the chargino and gluino 
masses are about 80 and 150 GeV, respectively \cite{PDG}. 
This means that the low energy values of $M_2$ and $M_3$ should 
not be smaller than these numbers. The one--loop RGEs tell us 
that the ratio of the gaugino mass $M_a$ $(a=1,2,3)$ at two 
different energy scales is the same as the ratio of the corresponding 
gauge coupling constants $\alpha_a$ at the same scales. 
Using the known evolution of the gauge coupling constants 
and the above experimental limits we obtain bounds on $M_2$ and 
$M_3$ at the high energy scale:
\be
M_2 |_{M_X} \simgt 100 \GeV
\,,
\qquad
M_3 |_{M_X} \simgt 50 \GeV
\,.
\ee
We do not expect that the actual gaugino mass parameters are much 
bigger than the above lower limits. Remember that in the models 
considered here the soft scalar masses are bigger by a factor 
$\cO(1/\epsilon)$ which can be at least $\cO(100)$. So, the soft 
scalar masses are already in the range of tens of TeV. They should 
not be bigger if supersymmetry is to cure the hierarchy problem.. 
Thus, we conclude that in this class of models the gaugino 
mass parameters at the high energy scale are of the order 
of the weak scale\footnote
{We do not expect $M_1$ to be much bigger than $M_2$ and $M_3$. 
In fact in many cases (for example for the universal gaugino 
masses at $M_X$) $M_1$ is the smallest gaugino mass at the weak scale.
}
\be
M_a |_{M_X} \approx \cO(M_Z)
\,.
\ee

Here we have to take a closer look at the Yukawa couplings.
It is well known that the  RGE of the top-quark Yukawa
coupling has an infra--red 
(quasi)--fixed point. The measured top quark mass and the analysis of 
the evolution of the bottom quark to the tau lepton mass ratio 
suggest that the actual top Yukawa coupling is not far from that fixed 
point value. The existence of such a fixed point is very important 
for the evolution of the soft scalar masses. We use the parameter $y$ 
to measure how close we are to the fixed point:
\be
y = \frac{Y}{Y_f}
\ee
where $Y = h^2_t / 4\pi$ ($h_t$ being the top Yukawa coupling) 
and $Y_f$ is its fixed point value. 
The experimental data are not precise enough to tell us how far 
exactly we are from the fixed point corresponding to $y=1$. 
We know however that the actual value of the parameter $y$ 
is not smaller than about 0.9 which we will use further as a typical 
value.

Now we can consider the RGEs of the soft 
scalar masses. They are of the form
\be
\frac{d}{dt} m_i^2 
=
- c_i Y \left( \mbar^2 + A^2 \right)
+ \ldots
\label{eq:RGEm}
\ee
where $\mbar^2 = m_{H_2}^2 + m_{U_3}^2 + m_{Q_3}^2$, 
$A$ is the trilinear soft term for the top quark, 
$c_{H_2}=3$, $c_{U_3}=2$, $c_{Q_3}=1$ and $c_i=0$ 
for other scalars. 
The dots stand for small contributions proportional to 
squares of the gaugino masses or to squares of Yukawa couplings 
other than that of the top quark 
(we assume here that $\tan\beta = v_2/v_1$ is not as large as
its maximal value of about $m_t/m_b$). 
We can see that only three soft scalar masses:
$m_{H_2}^2$, $m_{U_3}^2$ and $m_{Q_3}^2$ 
are substantially renormalized.  
All other low energy soft masses are very close  
to their initial values at $M_X$. 
The solution to (\ref{eq:RGEm}) for these three masses 
(taking into account also the RGE for $A$) is given by 
\cite{CCOPW}
\be
m_i^2
=
\left(m_i^2\right)_0 
- \frac{c_i}{6} \left[ y \mbar^2_0 + y(1-y)A_0^2 \right]
+ \ldots
\ee
where subscript 0 denotes the initial values at the high energy 
scale $M_X$. Summing up the three solutions we get
\be
\mbar^2
=
(1-y) \left( \mbar^2_0 - y A_0^2 \right)
\,.
\ee
The parameter $y$ is quite close to 1, thus, 
the sum of the squares of those 3 soft masses at the weak scale 
is much smaller than at the high energy scale. 
It can be even negative if $A_0$ is big enough. 
The two soft squark parameters, 
$m_{U_3}^2$ and $m_{Q_3}^2$, must be positive 
because they determine (up to a mixing) the masses of 
the left-- and right--handed top squarks.
On the other hand, the third mass, 
$m_{H_2}^2$, 
should be negative in order to trigger the radiative gauge 
symmetry breakdown. 
The renormalization is different for the three soft masses 
considered here. The negative contribution to the Higgs soft parameter, 
$m_{H_2}^2$, is 3 (1.5) times bigger than the corresponding 
contribution to $m_{Q_3}^2$ ($m_{U_3}^2$). 
Thus, the most natural solutions to the above constraints 
give at low energies:
\be
m_{H_2}^2 = -\cO(m_{3/2}^2)
\,,
\qquad
m_{U_3}^2 \,\,, m_{Q_3}^2 = \cO(m_{3/2}^2)
\ee
up to some coefficients of order unity 
(much bigger than $\epsilon$).
The important information for our analysis is that 
the absolute values of these soft terms are much bigger 
than the weak scale $M_Z$. 

There are two possible exceptions from this pattern but 
both require strong fine--tuning of the initial parameters. 
One of the squark soft parameters may be much smaller than 
the above typical value. In such a situation the mass of 
one of the top squarks can be as small as the weak scale 
(especially in the presence of a strong stop mixing). 
This requires a fine--tuning of the initial value 
of one of the soft squark parameters. 
In principle it is also possible that the absolute values 
of all three soft masses are of the order of $M_Z$ instead 
of $m_{3/2}$. This however may happen only if we fine--tune 
all three initial values to satisfy the condition 
$m_{H_2}^2 : m_{U_3}^2 : m_{Q_3}^2 \sim 3:2:1$ 
at high energy scale $M_X$.

Keeping in mind the possible exceptions we conclude that 
the most natural spectrum of the low energy soft SUSY  
breaking masses is the following: 
the square of the soft mass of $H_2$ doublet is negative 
with the absolute value of the order $\cO(M_Z^2/\epsilon^2)$; 
all other soft scalar masses are positive and of the same 
order of magnitude. 

Let us now consider the radiative breakdown of the electroweak 
gauge symmetry. The $Z$ boson mass is given by the equation
\be
M_Z^2
=
2 \frac{m_1^2 -\tan^2\beta m_2^2}{\tan^2\beta -1}
\ee
where $m_i^2 = m_{H_i}^2 +\mu^2$ and $\tan\beta=v_2/v_1$. 
{} From this formula we can calculate the value of 
the parameter $\mu$ describing the supersymmetric mixing 
of the two Higgs doublets:
\be
\mu^2
=
\frac{m_{H_1}^2 -\tan^2\beta m_{H_2}^2}{\tan^2\beta - 1}
-
\frac{1}{2}M_Z^2  \label{mu2}
\,.
\ee
The r.h.s.\ of the above expression is dominated by the first 
term\footnote
{This is true even if we 
fine--tune $m_{H_2}^2$ to small values of order $M_Z^2$. 
} 
which is of the order of $m_{3/2}^2$. Thus, we get 
\be
\mu = \cO\left(m_{3/2}\right) = \cO\left(M_Z/\epsilon\right) 
\,.  \label{mu-parameter}
\ee

The soft term $m_{H_2}^2$ is big and negative 
but the Higgs potential parameter $m_2^2$ is not. 
Using the formula for $\tan\beta$:
\be
\tan^2\beta
=
\frac{m_1^2 + \frac{1}{2}M_Z^2}{m_2^2 + \frac{1}{2}M_Z^2}
\ee
we obtain the bound
\be
m_2^2 > - \frac{1}{2} M_Z^2
\,.
\label{eq:m22}
\ee

Now we will use all the above informations to get the most 
characteristic features of the SUSY spectrum relevant for 
the calculation of the LSP relic abundance.
The lightest neutralino (LSP) is the mixture of the four neutral 
superpartners:
\be
\tilde{\chi} 
=
N_1 \tilde B + N_2 \tilde W^3 + N_3 \tilde H_1^0 + N_4 \tilde H_2^0
\,.
\ee
Without fine--tuning we have
\be
\mu = \cO\left(m_{3/2}\right) \gg M_1 \,, M_2 \,, M_Z 
\,.
\label{eq:mu}
\ee
In such limit the LSP is almost a pure gaugino. We have to consider 
two cases depending on the relative size of $M_1$ and $M_2$ 
parameters. 
On the other hand, the chargino mass matrix does not depend on $M_1$. 
Thus, in the limit (\ref{eq:mu}) the lighter chargino 
is almost pure gaugino with mass close to $M_2$. 
The masses and compositions of the LSP and the lighter chargino 
in  leading order in $M_Z/\mu \sim \epsilon$ expansion 
are given in table 1.

Usually $M_1$ is the smallest gaugino mass at low energies. 
In such a case the LSP is almost pure bino. We will concentrate 
on that possibility in the most part of our analysis. 
However, $M_2$ can be smaller for some non-universal gaugino 
masses at $M_X$. In this case the LSP is almost pure wino and 
has a mass very close to that of the lighter chargino. 
The coannihilation processes are very important in such a case. 
This situation has been considered in refs.\cite{MNY,CDG}. 
\begin{table}
\begin{center}
\begin{tabular}{|c|c|c|}\hline\hline
&  $M_1 < M_2$  &  $M_2 < M_1$  \\
\hline\hline
$m_{\tilde{\chi}}$  &  $M_1$  &  $M_2$  \\
\hline
$N_1$  &  1  &  0  \\
\hline
$N_2$  &  0  &  1  \\
\hline
$N_3$  &  $\sin\theta_W\sin\beta\frac{M_Z}{\mu}$  
       &  $-\cos\theta_W\sin\beta\frac{M_Z}{\mu}$  \\
\hline
$N_4$  &  $-\sin\theta_W\cos\beta\frac{M_Z}{\mu}$
       &  $\cos\theta_W\cos\beta\frac{M_Z}{\mu}$  \\
\hline\hline
$m_{\tilde{\chi}^+}$  &  
\multicolumn{2}{|c|}{$M_2$}  \\
\hline
$\phi_U$  &  
\multicolumn{2}{|c|}{$-\sqrt{2}\cos\theta_W\sin\beta\frac{M_Z}{\mu}$}  \\
\hline
$\phi_V$  &
\multicolumn{2}{|c|}{$-\sqrt{2}\cos\theta_W\cos\beta\frac{M_Z}{\mu}$}  \\
\hline\hline
\end{tabular}
\caption{The lightest neutralino and chargino 
masses and compositions in the leading order in $M_Z/\mu$.}
\end{center}
\end{table}

Let us now have a closer look at the Higgs spectrum. 
The pseudoscalar mass square 
$m_A^2$ is approximately equal to the sum  $m_1^2 + m_2^2$. 
Using eq.\ (\ref{eq:m22}) we find that the pseudoscalar mass 
is of the order of $m_{3/2}$. 
Masses of $H^0$ and $H^\pm$ are also of similar size:
\be
m_A \approx m_{H^0} \approx m_{H^\pm} = \cO\left(m_{3/2}\right)
\,.
\label{eq:mAmH}
\ee
Thus, $h^0$ is the only light Higgs scalar in the spectrum.
In the relic abundance calculation we will need 
the Higgs mixing angle $\alpha$. 
In the limit of very heavy $A$ and $H^0$ bosons 
it is determined by equations 
\be
\sin 2\alpha 
\approx 
-\sin 2\beta \frac{m_H^2 + m_h^2}{m_H^2 - m_h^2} 
\approx
-\sin 2\beta
\ee
\be
\cos 2\alpha 
\approx 
-\cos 2\beta \frac{m_A^2 - M_Z^2}{m_H^2 - m_h^2} 
\approx
-\cos 2\beta
\ee
implying
\be
\alpha \approx \beta +\frac{\pi}{2} .
\ee

%
%
\section{\bf Relic abundance of LSP in the weakly coupled case}

In SUSY models with $R$-parity invariance, the lightest SUSY
particle is stable and can constitute a significant portion of
the mass of the universe \cite{LSP-DM}.  On the other hand, we have
the following upper bound on the mass density of the LSP
\begin{eqnarray}
\Omega_{\tilde \chi} h^2 \simlt 1
\label{LSP-cond}
\end{eqnarray}
in order not to overclose the universe.
Here $\Omega_{\tilde \chi}$ is the mass density of the LSP relative to 
the critical density $\rho_c \approx 1.88 \cdot 10^{-29}$g/cm$^3$ 
and $h$ is the Hubble constant in units of $100$ km/s/Mpc.

In this section, we shall argue that, in most of the parameter space
allowed by the gaugino condensation scenario in the weakly
coupled case, the relic abundance of the LSP becomes too large,
resulting in the overclosure of the universe.

The relic abundance $\Omega_{\chi} h^2$ of the lightest neutralino 
is given by
\begin{eqnarray}
\Omega_{\tilde \chi} h^2 = {1.07 \times 10^9 {\mbox{GeV}}^{-1} x_F
\over \sqrt{g_*} m_{Pl} \overline{\sigma_A v}},
\label{RA-formula}
\end{eqnarray}
where $m_{Pl}=1.2 \times 10^{19}$ GeV, $x_F$ is defined as 
$x_F \equiv m_{\tilde{\chi}}/T_F$
in terms of the freeze-out temperature $T_F$, and
$g_*$ is the effective number of relativistic degrees of
freedom at $T_F$.
The typical values of these parameters are $15 \simlt x_F \simlt 30$
and $8 \simlt \sqrt{g_*} \simlt 10$.
$\overline{\sigma_A v}$ is the thermal average of the 
 annihilation cross section $\sigma_A $ times
the relative velocity $v$ of the $\tilde{\chi}
\tilde{\chi}$
pair in its center-of-mass frame, and 
\begin{equation}
  \overline{\sigma_A v}\equiv 
   \frac{\int^{T_{F}}_{T_0}  \sigma_A v dT/m}
     {\int^{T_{F}}_{T_0} dT/m}
  = x_F \int^{T_{F}}_{T_0}  \sigma_A v dT/m
\end{equation}
with $T_0 \ll T_F$.
When the freeze-out of $\tilde{\chi}$ occurs, 
the relative velocity is estimated to be 
$v^2 \sim 6/x_F = \cO(0.2 \sim 0.4)$.
Hereafter we use $v^2 = \cO(0.2)$.

Let us now estimate the cross section $\sigma_A v$ and
the relic abundance $\Omega_{\chi} h^2$. 
(The general formulae of the amplitudes for 
possible annihilation processes are given in \cite{DN}.)
For the moment we focus on the 
case where $|M_1|$ is smaller than $|M_2|$, which happens e.g.\ 
when the gaugino masses are universal at the string scale.
We do not consider the annihilation processes whose final
states include the scalar  bosons  $H^0$, $A$ and $H^{\pm}$ because 
they are kinematically forbidden (see eq.~(\ref{eq:mAmH})).

In the following, we ignore possible interference between various
channels. 
This simplification does not change our conclusions 
because usually only one channel dominates the cross section. 

\begin{enumerate}

\item{$\tilde{\chi} \tilde{\chi} \to f \bar{f}$}

When $\tilde{\chi}$ is lighter than $W$-boson,
the only final states allowed by kinematics are quark 
and lepton pairs $f\bar{f}$ (with $f \neq t$).

As we argued in the previous section, the lightest neutralinos $\tilde
\chi$ are usually gaugino--like. Then one would expect that they 
annihilate into fermion pairs mainly through $t$-channel
sfermion exchange.  However, in the case at hand, 
the exchanged sfermions are very 
heavy and the corresponding amplitude becomes small. Indeed
the magnitude of $\sigma_A v$ is proportional to
\begin{eqnarray}
\sigma_A v &\propto& {1 \over s}
\left({m_f m_{\tilde{\chi}} \over m_{\tilde{f}}^2}\right)^2 
~~~~~~~~~(s{\mbox{-wave}}) ,
\label{sigma-ffs}\\
&\propto& {v^2 \over s}\left({m_{\tilde{\chi}} \over
m_{\tilde{f}}}\right)^4
~~~~~~~~~~~~(p{\mbox{-wave}})
\label{sigma-ffp}
\end{eqnarray}
in the limit $m_{\tilde{\chi}} \ll m_{\tilde{f}}$ 
(here $s$ is  the center-of-mass energy squared and  
$m_f$ represents the mass of the fermion in the final state).
The $p$-wave contribution is dominant because 
$m_f < m_{\tilde{\chi}}$.
The relic abundance is estimated as 
\begin{eqnarray}
\Omega_{\tilde{\chi}} h^2 
&\sim& 
 4 \times \left({m_{\tilde{f}}^2 \over 
(1 {\mbox{TeV}})m_{\tilde{\chi}}}\right)^2 
\nonumber \\
&\sim& 10^{-2} \epsilon^{-4} 
     \left(\frac{m_{\tilde \chi}}{50 \mbox{GeV}} \right)^2,
\label{RA-ff}
\end{eqnarray}
where $\epsilon \sim m_{\tilde \chi}/m_{\tilde f}$ was used.
Recalling that $\epsilon$ is a small number in the gaugino
condensation scenario in the weak coupling case, typically $\epsilon
\simlt1/100$, one finds from eq.~(\ref{RA-ff}) that
$\Omega_{\tilde{\chi}} h^2$ becomes much larger than unity.  In the
following we will explore whether the LSP 
can effectively annihilate via other channels 
to give a cosmologically acceptable level.

The magnitudes of the cross sections via the $s$-channel exchange of
$A$ and $H^0$ are estimated as
\begin{eqnarray}
\sigma_A v &\propto& {1 \over s}
\left({m_{\tilde{\chi}}^2 \over 4m_{\tilde{\chi}}^2-m_A^2}\right)^2
\left({m_f \over \mu}\right)^2  
\end{eqnarray}
and
\begin{eqnarray}
\sigma_A v &\propto& {v^2 \over s}
\left({m_{\tilde{\chi}}^2 \over 4m_{\tilde{\chi}}^2-m_{H^0}^2}\right)^2
\left({m_f \over \mu}\right)^2  ,
\label{sigma-ff2}
\end{eqnarray}
respectively.
These processes have smaller cross sections than the $t$-channel
sfermion exchange 
because $\sigma_A v$ includes a suppression factor $(m_f/\mu)^2$ in
addition 
to
${m_{\tilde{\chi}}^4/(4m_{\tilde{\chi}}^2-m_{A(H^{0})}^2})^2=
\cO(\epsilon^4)$.
The factor $(m_f/\mu)^2$ originates from  the coupling
of $\tilde{\chi}$ and $A(H^0)$:
$g_{\tilde{\chi}\tilde{\chi}A} =(g/2)\tan\theta_W \sin\theta_W (M_Z/\mu)$,
$g_{\tilde{\chi}\tilde{\chi}H^0} =(g/2)\tan\theta_W \sin\theta_W (M_Z/\mu)
\cos 2\beta$ and 
that of $f$, $\bar{f}$ and $A(H^0)$,
$g_{f\bar{f}A(H^0)} \propto (g/2)(m_f/M_W)\tan\beta$.
Resonance enhancement is not possible 
because $m_{A(H^0)} \gg m_{\tilde{\chi}}$.

In the same way,  the cross section through the $s$-channel 
exchange of the $h^0$ boson can be estimated to be 
\begin{eqnarray}
\sigma_A v &\propto& {v^2 \over s}
\left({m_{\tilde{\chi}}^2 \over 
4m_{\tilde{\chi}}^2-m_{h^0}^2 + i\Gamma_{h^0} m_{h^0}}\right)^2
\left({m_f \over \mu}\right)^2  
\label{sigma-ff2'}
\end{eqnarray}
where $\Gamma_{h^0}$ is the decay width of the $h^0$-boson.
We obtain the relic abundance of the order 
\begin{eqnarray}
\Omega_{\tilde{\chi}} h^2 \simgt 2 \times
 \left({\mu m_{h^0}^2 \over (1 {\mbox{TeV}})m_f m_{\tilde{\chi}}}\right)^2 
\label{RA-ff2'}
\end{eqnarray}
if $m_{\tilde \chi}$ is not too close to half of $m_{h^0}$. One readily
finds that $\Omega_{\tilde{\chi}} h^2$ is unacceptably large.  For
instance,  for $m_{\tilde{\chi}} = \epsilon \mu$, $m_{h^0} = 100$GeV,
$m_f = 5$GeV we get $\Omega_{\tilde{\chi}} h^2 \simgt \cO(8 \times
\epsilon^{-2})$, which is always larger than unity. 

Let us now consider the case where a resonance enhancement occurs,
i.e., $m_{\tilde{\chi}} \sim m_{h^0}/2$.  Careful
treatment near a pole was discussed in 
Refs.~\cite{GriestSeckel,GondoloGelmini}. Since the
decay width of the Higgs boson is very narrow $\Gamma_{h^0} / m_{h^0}
\sim 2 \times 10^{-5}$, one may approximate the Higgs boson propagator
by a delta function. Following the argument of
Ref.~\cite{GondoloGelmini}, we find
\begin{equation}
\overline{\sigma_A v} =
x_F \frac{16 \pi \omega}{m^2_{\tilde{\chi}}} \pi \gamma_R 
    \mbox{erfc}(x_F^{1/2} \epsilon_R^{1/2}) 
     b_R(\epsilon_R) \theta(\epsilon_R)
\label{eq:cross-section:resonance}
\end{equation}
where
\begin{equation}
   \gamma_R \approx \frac{\Gamma_{h^0}}{m_{h^0}},~~~
   \epsilon_R=\frac{m_{h^{0}}^2-4 m_{\tilde \chi}^2}{4 m_{\tilde \chi}^2},
   ~~~\omega=\frac{1}{4}
\end{equation}
and 
\begin{equation}
  b_R(\epsilon_R)\approx 
  \frac{B(h^0 \rightarrow \tilde \chi \tilde \chi)}{\epsilon_R^{1/2}}.
\end{equation}  
The branching 
ratio $B(h^0 \rightarrow \tilde \chi \tilde \chi)$ is evaluated as
\begin{equation}
  B(h^0 \rightarrow \tilde \chi \tilde \chi)
  =\frac{c g^{\prime 2}/32 \pi (M_Z/\mu)^2 \beta^3 m_{h^0}}{\Gamma_{h^0}}
\end{equation}
with the velocity of the neutralino $\beta \approx 2 \epsilon_R^{1/2}$.
The constant $c\sim \cO(1)$ depends on $\tan \beta$. 
Plugging the above formulae into 
eq.~(\ref{eq:cross-section:resonance}), we find
\begin{equation}
\overline{\sigma_A v } 
=\frac{32\pi^2 }{m_{\tilde \chi}^2} \times c \frac{g^{\prime 2}}{4\pi}
\left( \frac{M_Z}{\mu} \right)^{2} 
\mbox{erfc}(x_F^{1/2} \epsilon_R^{1/2}) \epsilon_R x_F
\,.
\end{equation}
The function erfc$(x_F^{1/2} \epsilon_R^{1/2}) \epsilon_R x_F$ 
takes its maximum value of about 0.16 for 
$\epsilon_R x_F \approx 0.64$.  With the Higgs boson
mass giving the value of $\epsilon_R \approx 0.64 x_F^{-1}$, 
the annihilation cross section is maximal:
\begin{equation}
  \overline{\sigma_A v}|_{max} \sim \frac{0.5 c}{m_{\tilde \chi}^2}
 \left( \frac{M_Z}{\mu} \right)^{2}
\end{equation}
which in turn gives the minimum of the relic abundance
\begin{equation}
 \Omega_{\tilde \chi}h^2 |_{min}
 \sim 10^{-6} \epsilon^{-2} 
    \left(\frac{m_{\tilde \chi}}{50 \mbox{GeV}} \right)^{2}
\,.
\end{equation}
Thus, the relic abundance can be optimized to be smaller than 
unity by appropriately choosing the masses of the Higgs boson 
and the neutralino even for a typical value of $\epsilon =\cO(1/100)$.  
Note that the function 
$x$erfc$x^{1/2}$ decreases rapidly as $x$ deviates from 0.64, and thus a
small change of the Higgs mass increases the relic abundance 
drastically. Indeed if one increases the Higgs mass by 10\% from the
optimal value, the relic abundance increases by more than one order of 
magnitude. The increase is even faster when one decreases the Higgs mass.
Thus, we conclude that the resonance enhancement by $h^0$ 
exchange can reduce the relic abundance to a cosmologically viable 
level only for a very small range
of the Higgs boson mass close to $2m_{\tilde\chi}$.

The last contribution to the 
${\tilde{\chi}}{\tilde{\chi}}\rightarrow f{\bar f}$ 
annihilation comes from the $s$-channel exchange of the $Z$ boson. 
The corresponding cross section is proportional to
\begin{eqnarray}
\sigma_A v &\propto& {1 \over s} 
\left({m_f m_{\tilde{\chi}} \over \mu^2}\right)^2
~~~~~~~(s{\mbox{-wave}}) , \\
&\propto& {v^2 \over s} 
\left({M_Z \over \mu}\right)^4 
~~~~~~~~~(p{\mbox{-wave}}) 
\label{sigma-ff3}
\end{eqnarray}
as far as we are not close to the $Z$-resonance. The  suppression factors  
$({m_f m_{\tilde{\chi}} /\mu^2})^2$ and $({M_Z / \mu})^4$
make the cross section too small. So the question is again whether
the $Z$-resonance can sufficiently  enhance the cross section.  Compared to 
the previous case of the $h^0$ resonance, the $Z$ decay width is rather
broad 
($\Gamma_Z/M_Z \sim 0.027)$ and we should
use a different approximation to evaluate the annihilation cross section.
As a crude estimate of the maximal cross section, one may replace the 
$Z$-boson propagator by \cite{GriestSeckel}
\begin{eqnarray}
{1 \over s - M_Z^2 + i \Gamma_Z M_Z} \sim {1 \over M_Z^2 v^2}. 
\label{Z-propagator}
\end{eqnarray}
Then one finds the relic abundance to be
\begin{eqnarray}
\Omega_{\tilde{\chi}} h^2 \sim 7.2 \times 
\left({\mu \over 1 {\mbox{TeV}}}\right)^2 
\left({v \mu \over m_{\tilde{\chi}}}\right)^2 
\sim \left({\mu \over 1 {\mbox{TeV}}}\right)^2 
\left({ \mu \over m_{\tilde{\chi}}}\right)^2. 
\label{RA-Z-ff}
\end{eqnarray}
Thus, it is too large when $\mu \gg m_{\tilde \chi}$ which is the case in
this scenario.

\item{$\tilde{\chi} \tilde{\chi} \to W^+ W^-, ~~Z Z$}

When $\tilde{\chi}$ is heavier than the $Z$ boson, $W^+ W^-$ and $Z Z$
final states are kinematically allowed 
(only $W^+ W^-$ if $M_W < m_{\tilde \chi} < M_Z$). 
The dominant contribution to these processes is the 
$s$-channel exchange of the lightest Higgs boson $h^0$.
The magnitude of $\sigma_A v$ is estimated as
\begin{eqnarray}
\sigma_A v &\propto&
 {v^2 \over s} \left({M_Z \over \mu}\right)^2 
\left({M_Vm_{\tilde{\chi}} \over 4m_{\tilde{\chi}}^2-m_{h^0}^2}\right)^2
\left({m_{\tilde{\chi}} \over M_V}\right)^4
\nonumber\\
&\sim&
 {v^2 \over s} \left({m_{\tilde{\chi}} \over \mu}\right)^2 
\left({m_{\tilde{\chi}}^2 \over 4m_{\tilde{\chi}}^2-m_{h^0}^2}\right)^2
\left({M_Z \over M_V}\right)  ^2
\label{sigma-VV}
\end{eqnarray}
where $M_V$ denotes gauge boson masses $(V = W^{\pm}, Z)$.
The relic abundance is approximately given by 
\begin{eqnarray}
\Omega_{\tilde{\chi}} h^2 \sim  5.4 ({\mbox{or}}~ 20) \times 
\left({\mu \over 1 {\mbox{TeV}}}\right)^2 
\left({4m_{\tilde{\chi}}^2-m_{h^0}^2 \over m_{\tilde{\chi}}^2}\right)^2 
\label{RA-VV}
\end{eqnarray}
for $\tilde{\chi} \tilde{\chi} \to W^+ W^-$ (or $Z Z$).
The factor $(M_Z/\mu)^2$ stems from the coupling 
among $\tilde{\chi}$, $\tilde{\chi}$ and $h^0$, 
$g_{\tilde{\chi}\tilde{\chi} h^0} = g\tan\theta_W\sin\theta_W(M_Z/\mu)
(\sin2\beta/2)$ and 
the factor $(m_{\tilde{\chi}}/M_V)^4$ reflects the enhancement of the
amplitude when the gauge bosons in the final state have longitudinal
polarization. 
Again one finds that the 
annihilation cross section is not big enough to reduce the
relic abundance to an acceptable level.  Note that when $\tilde \chi$ is 
heavier than the $W$ boson,  the resonance enhancement 
$m_{\tilde{\chi}} \sim m_{h^0}/2$ does not occur 
because there is an upper bound on $m_{h^0}$ 
in the MSSM which is much below $2 M_W$
\cite{h-mass}.

Let us now briefly mention other channels.
The cross section for the process $\tilde{\chi}\tilde{\chi} \to W^+W^-$
via the $s$-channel exchange of the $Z$ boson is as tiny as 
$\sigma_A v \sim s^{-1}(M_W m_{\tilde{\chi}}/\mu^2)^2
\sim \cO(s^{-1}\epsilon^4)$.
This is analogous to the corresponding channel in the process 
$\tilde{\chi}\tilde{\chi} \to f\bar{f}$.
The cross section includes the suppression factor $(M_Z/\mu)^4$ or 
$(M_Z/\mu)^6$
for the process $\tilde{\chi}\tilde{\chi} \to W^+W^-$
via the $t$-channel exchange of the lighter chargino
or the heavier one, respectively, and hence this process is 
also much suppressed. 
This is due to the facts that the coupling among 
$\tilde{\chi}$, $\tilde{\chi}^{\pm}$ and $W^{\mp}$ is proportional to 
$M_Z/\mu$ and the propagator of the heavier chargino behaves like
$1/\mu^2$. 
In the same way, the cross section includes the suppression factor 
$(M_Z/\mu)^4$ and $(M_Z/\mu)^6$
for the process $\tilde{\chi}\tilde{\chi} \to ZZ$ 
via the $t$-channel exchange of lighter neutralinos 
$\tilde{\chi}_{1(2)}^0$
and heavier ones, respectively.

\item{$\tilde{\chi} \tilde{\chi} \to h^0 h^0$}

When $\tilde{\chi}$ is heavier than $h^0$, the $h^0 h^0$ channel is  open.
The dominant contribution for this process is also the 
$s$-channel exchange of $h^0$.
The magnitude of $\sigma_A v$ is estimated as
\begin{eqnarray}
\sigma_A v \propto {v^2 \over s} \left({M_Z \over \mu}\right)^2 
\left({M_Z m_{\tilde{\chi}} \over 4m_{\tilde{\chi}}^2-m_{h^0}^2}\right)^2
\label{sigma-hh}
\end{eqnarray}
(the suppression factor $(M_Z/\mu)^2$ originates from the coupling
among $\tilde{\chi}$, $\tilde{\chi}$ and $h^0$) and the relic 
abundance is given by 
\begin{eqnarray}
\Omega_{\tilde{\chi}} h^2 \sim 27 \times 
\left({\mu \over 1 {\mbox{TeV}}}\right)^2 
\left({4m_{\tilde{\chi}}^2-m_{h^0}^2 \over M_Z^2}\right)^2.
\label{RA-hh}
\end{eqnarray}
This process also induces too large relic abundance
(observe that the extra factor of $\cO(10)$ is generated
from $(4m_{\tilde{\chi}}^2-m_{h^0}^2)^2/M_Z^4$) and 
the resonance enhancement at 
$m_{\tilde{\chi}} \sim m_{h^0}/2$ is not possible 
because $m_{\tilde{\chi}} \simgt m_{h^0}$.

The cross section for  $\tilde{\chi}\tilde{\chi} \to h^0h^0$ 
via the $t$-channel exchange of lighter neutralinos 
$\tilde{\chi}_{1(2)}^0$ 
includes a suppression factor 
$(M_Z/\mu)^4$
 and thus is small.
A similar process through heavy neutralino exchange 
gives a relatively large contribution 
because the couplings among $\tilde{\chi}_{1(2)}^0$,
$\tilde{\chi}_{3(4)}^0$
and $h$ have no suppression 
(e.g.\ $g_{\tilde{\chi}_1 \tilde{\chi}_{3} h^0} = g\tan\theta_W \sin\alpha
/2\sqrt{2}$). 
But even in this case we obtain the relic abundance which is too large:
\begin{eqnarray}
\Omega_{\tilde{\chi}} h^2 &\simgt& 
9.6 \times \left({\mu \over 1 {\mbox{TeV}}}\right)^2 
\,.
\label{RA-hh2}
\end{eqnarray}

\item{$\tilde{\chi} \tilde{\chi} \to Z h^0$}

Because the coupling 
$g_{\tilde{\chi}\tilde{\chi}Z} \propto (M_Z/\mu)^2$ is small, the
$s$-channel 
exchange of $Z$ is not effective. The $t$-channel exchange of the 
neutralinos yields the cross section $\propto s^{-1} (m_{\tilde
\chi}/\mu)^2
(m_Z/\mu)^2$, which is again too small. Finally the $s$-channel $A$
exchange is
even more suppressed as long as $m_A=\cO(m_{3/2})$.

\end{enumerate}

To summarize, we find  it almost impossible for the bino-like LSP to
  satisfy the condition $\Omega_{\tilde{\chi}} h^2 \simlt 1$  in
  models with mass spectra characterized by eqs.~(\ref{scalar-mass}),
(\ref{gaugino-mass}), (\ref{mu-parameter}) and (\ref{eq:mAmH}). 
The only exceptional case is when the LSP mass is fine tuned to nearly
half of the mass of the Higgs boson $h^0$ with precision better 
than about 10\%.  The resonance is effective 
only in a small region of  parameter space.

Before ending this section, we would like to explore other possible
ways to avoid the overclosure problem discussed above.

One way is to allow for a certain amount of fine-tuning among the
soft-breaking parameters. There are several possibilities:
\begin{itemize}
\item
One may try to adjust
$\mu$ at $\cO(100)$ GeV, by fine--tuning the parameters in the r.h.s.\ of
eq.~(\ref{mu2}). Then $m_{Z}/\mu$, $m_{\tilde \chi}/\mu$ etc.\ are no
longer suppression factors in the annihilation cross section and thus
one can obtain $\Omega_{\tilde \chi} h^2$ less than unity.  
\item
Fine--tuning may lead to the situation that $m_A$ or $m_{H^0}$ is small and
comparable to $m_{\tilde{\chi}}$ in addition to the very small $\mu$
parameter.  Then 
the $s$-channel exchange of $A$ (or $H^0$) can be a dominant contribution
in the
process $\tilde{\chi} \tilde{\chi} \to Zh^0$ with the cross section
\begin{eqnarray}
\sigma_A v &\propto& {1 \over s} \left({m_{\tilde{\chi}} \over
\mu}\right)^2
\left({M_Z^2 \over m_A^2 + M_Z^2}\right)^2 
\left({m_{\tilde{\chi}}^2 \over 
4m_{\tilde{\chi}}^2-m_A^2 + i\Gamma_A M_A}\right)^2 
\label{sigma-Zh2}
\end{eqnarray}
where $\Gamma_A$ is the decay width of the $A$-boson.
Such a process can lead to a realistic amount of relic abundance
near the pole $m_{\tilde{\chi}} \sim m_A/2$.
\item
Another possibility appears when $m_{\tilde{\chi}}$ is 
larger than the top quark mass  one of the stop masses 
$m_{{\tilde{t}}_1}$
may be fine--tuned to be much smaller than the gravitino mass. 
Using eq.\ (\ref{RA-ff}), we can find that the condition (\ref{LSP-cond})
is fulfilled if $m_{{\tilde{t}}_1}^2/m_{\tilde{\chi}} \simlt 1$TeV.
\end{itemize}
Though these loop-holes are possible, they require severe fine--tunings of
the parameters and are very unlikely. 

Finally one can obtain a small relic abundance of the LSP when
one considers the case with $|M_2|$ smaller than $|M_1|$.  
In such a situation, the LSP is dominantly the 
neutral component of wino $\tilde{w}_3$ and the relic density of
$\tilde{\chi}$ is reduced by a co-annihilation process, $\tilde{\chi}
\tilde{\chi}^{\pm} \to \gamma W^{\pm}, f\bar{f}'$, so that the
condition (\ref{LSP-cond}) is satisfied \cite{MNY,CDG}.  
This situation could be realized in some special cases. One example 
is in string models with non-universal gaugino masses at the 
string scale $M_{st}$. To illustrate this, let us consider
 a string model with the standard model 
gauge group and the MSSM particle contents. 
Here $M_{st}$ is defined by 
$M_{st} \equiv 0.527 \times g_{st} \times 10^{18}$ GeV
\cite{Mst} where $g_{st}$ is a gauge coupling at $M_{st}$ at tree level
which is given by $g_{st} = \langle Re S \rangle^{-1/2}$.
By the use of the structure constants $\alpha_a$ at $M_{st}$ given by 
$\alpha_a(M_{st}) = (4\pi \langle Re (S + \epsilon_a T) \rangle)^{-1}$,
the gaugino masses $M_a$ at $M_{st}$ are given by
\begin{eqnarray}
M_a(M_{st}) &=& 8\pi \epsilon_a \langle F_T \rangle \alpha_a(M_{st})
\label{Ma}
\end{eqnarray}
in the moduli dominant SUSY breaking scenario.
Here $\epsilon_a$'s are small quantities stemming from one loop corrections
to gauge kinetic functions.
The mass ratio between $SU(2)_L$ and $U(1)_Y$ gauginos 
evaluated at the weak scale of order $M_Z$ 
is given by
\begin{eqnarray}
{M_2(M_Z) \over M_1(M_Z)} &=&  
   {\alpha_2(M_Z) (\alpha_{st}^{-1} - \alpha_2(M_{st})^{-1})
   \over \alpha_1(M_Z) (\alpha_{st}^{-1} - \alpha_1(M_{st})^{-1})} .
\label{Ma-ratio2}
\end{eqnarray}
If $\alpha_{st}^{-1} > \alpha(M_X)^{-1} 
+ {23 \over 10\pi} \times (3.43 - \ln g_{st})$,
the $U(1)_Y$ gaugino is lighter than $SU(2)_L$ ones.
Here we have used the renormalization group flow  
of the gauge couplings based on the MSSM 
and $\alpha(M_X) \equiv \alpha_1(M_X) = \alpha_2(M_X)$.
Hence we have the same problem of large relic abundance of the LSP
as that in the string model discussed just before.
On the other hand, if $\alpha_{st}^{-1} < \alpha(M_X)^{-1} 
+ {23 \over 10\pi} \times (3.43 - \ln g_{st})$,
the $U(1)_Y$ gaugino is heavier than the $SU(2)_L$ ones and
the small relic abundance can be obtained through
the co--annihilation process, 
$\tilde{\chi} \tilde{\chi}^{\pm} \to \gamma W^{\pm}, f\bar{f}'$.

%
%
\section{Strongly coupled heterotic string theory}

We now turn to the strongly coupled case. Ho\v{r}ava and Witten \cite{HW}
showed
that the strong coupling limit of the $E_8 \times E_8'$ heterotic
string theory is described by the M-theory compactified on $S^1/Z_2$ 
(heterotic M-theory).  At each boundary of the
$S^1/Z_2$, an $E_8$ super Yang-Mills theory must be attached, due to
anomaly cancellation. Interestingly this heterotic M-theory allows one to
identify the compactification scale of the extra six dimensional space 
$X_6$ with the GUT scale of about $3 \times 10^{16}$ GeV 
inferred by the electroweak precision measurements \cite{C-unif,BD}.

The standard embedding of a Calabi-Yau manifold breaks one of the
$E_8$ gauge groups to $E_6$, leaving the other $E'_8$ unbroken.
Particles of the supersymmetric standard model live on the
$E_6$ wall. One can analyze properties of this theory at low energies
using effective four-dimensional supergravity \cite{NOY}. For this
purpose, one should appropriately integrate over the coordinates of the
internal space $X_6 \times S^1/Z_2$ to define fields in the
four-dimensional effective theory. It was shown 
\cite{NOY} that in the 
leading order it is basically enough to average over $S^1/Z_2$. 
This averaging procedure allows one also 
to derive the K\"{a}hler potential, superpotential and
gauge kinetic functions of the four-dimensional theory from the
Ho\v{r}ava-Witten Lagrangian of the M-theory. For example, one finds
the gauge kinetic functions to be
\begin{equation}
   f_{6,8}=S \pm \alpha T, \label{eq:gkf-strong}
\end{equation}
with $S,$ $T$ defined by appropriately averaging over the eleventh
dimensional interval. Here $\alpha$ is a numerical constant of order
unity, which is expressed as an integral over the Calabi-Yau manifold.
The second term in (\ref{eq:gkf-strong}) originates from gauge
anomaly cancellation, thus is a  next-to-leading order correction. 
The form of the kinetic functions is similar to the weak coupling
case, the difference being that, in the strong coupling regime, the
vacuum expectation value of $\alpha T$ is comparable to that of $S$.

Now suppose that gaugino condensation occurs at the $E'_8$ wall. It was
argued \cite{Horava} that supersymmetry has to be broken in this
case. This has been shown explicitely in \cite{NOY} by applying the 
averaging proceedure to intregrate out the heavy Kaluza-Klein
modes. The method allows one to identify which
auxiliary component of a scalar superfield has a vacuum expectation
value and is responsible for supersymmetry breaking. It turns out that
the gravitino mass is related to the gaugino condensation scale
$\Lambda$ as $ m_{3/2} \sim \Lambda^3/m_{Pl}^2$.  The scalar masses
are model dependent, but are of the order of the gravitino mass. 
So far, these
properties are similar to the case of the weakly coupled theory. A
crucial difference between the strong and weak coupling cases 
can be seen by investigating the gaugino masses. 
The large next-to-leading
order correction of the gauge kinetic function in the strong coupling
case makes the gaugino mass comparable to the gravitino mass. In contrast
to the weakly coupled case, gaugino masses are 
generically of the same size
as the scalar masses.

This sparticle mass spectrum leads to different phenomenological
and/or cosmological consequences from those of the weakly coupled 
theory. In the weak coupling case, the gaugino condensation scenario
gives the sparticle mass spectrum with the gaugino masses of about
two orders of magnitude smaller than the scalar masses. With this mass
spectrum, as was intensively studied in the previous section, the
relic abundance of the neutralino-LSP would be too large in most of
the parameter space. On the other hand, in the strong coupling regime,
the scalar masses and the gaugino masses are comparable, both of which
are assumed to be at the electroweak scale.  Then the annihilation of
the neutralinos through, for example, the $t$-channel sfermion exchange
is much more effective, reducing the relic abundance substantially.

A precise prediction of the relic abundance is very model
dependent\footnote{See \cite{BKL} for an analysis on the relic
  abundance in a special case.}
given the model dependence of the size of the
scalar masses discussed previously.  
We therefore do not need to go into much detail here.  
Rather we expect that,
independent of models and fine tunings of
parameters as was needed in the weakly coupled case,  
we can easily realize a situation where
the relic abundance does not exceed the closure limit. A
situation where $\Omega_{\tilde \chi} h^2$ is of the order of 0.1 and the
neutralino constitutes a dark matter of the universe is thus
to be expected.\footnote{In the
  M-theory regime, there may appear an axion field whose decay
  constant is as large as $\sim 10^{16}$ GeV \cite{M-axion}.  The
  coherent oscillation of such an axion would overclose the
  universe. To cure this, one would invoke entropy production after
  the axion's oscillation begins. The entropy production may change
  our arguments of the relic abundance of the LSPs
  \cite{entropy}. However, since the whole structure of the
  non-perturbative effects to the axion potential in the heterotic
  M-theory is unclear at this moment, we discard the possibility of
  the appearance of the M-theory axion in this paper and restrict
  ourselves to a standard thermal history of the universe.}

%
%
\section{Conclusions}

In this paper, we have studied the question of the relic abundance of
a stable neutralino LSP in  heterotic string theory, with
supersymmetry broken by  hidden-sector gaugino condensates. In
the weakly coupled regime, the gaugino condensation scenario predicts
small gaugino masses in the observable sector, much smaller than
scalar masses and the gravitino mass. Furthermore, the renormalization
group analysis and the requirement of the correct electroweak gauge
symmetry breakdown shows that the masses of the Higgs bosons 
(with exception of the lightest one) 
as well as the supersymmetric Higgs mixing parameter, $\mu$,
become as large as the gravitino mass. 
This could only be avoided  through a strong fine
tuning among the soft masses. In most of parameter space
the relic abundance of the neutralino is too
large to be cosmologically consistent.  
We have identified exceptional
cases where the relic abundance becomes acceptably small. They require
fine tuning of the parameters, or the assumption that the $SU(2)_L$ gaugino
mass $M_2$ is smaller than the $U(1)_Y$ gaugino mass $M_1$. Though
possible, these cases seem to be unlikely.  Thus we conclude that a
realistic relic abundance is difficult to achieve in the framework of 
the weakly coupled heterotic string. 

This problem is easily overcome when one considers the strongly
coupled regime of the heterotic string theory (heterotic M-theory). In 
this case, the gaugino masses become comparable to the scalar
masses. With this mass spectrum, we can easily realize situations
where the neutralino relic abundance is within the closure limit,
consistent with the cosmological observations.

%
%
\section*{Acknowledgments}

This work was supported by
the European Commission programs ERBFMRX--CT96--0045 and CT96--0090
and by a grant from Deutsche Forschungsgemeinschaft SFB--375--95.
The work of M.O. was partially supported by
the Polish State Committee for Scientific Research grant 2 P03B 040 12.
The work of M.Y. was partially supported by
the Grant--in--Aid for Scientific Research from the Ministry of
Education, Science and Culture of Japan No.\ 09640333.
M.O. and M.Y. thank Physikalisches Institut, Universit\"at Bonn for 
kind hospitality during their stay.

%
%


\end{document}